\documentclass{article}
\usepackage{arxiv}

\usepackage[T1]{fontenc}
\usepackage[utf8]{inputenc}
\usepackage{csquotes}

\usepackage[sort&compress,square,numbers]{natbib}
\usepackage{mathptmx}
\usepackage{array}
\usepackage{amsmath}
\usepackage[colorlinks=true]{hyperref}
\usepackage{xspace}
\usepackage{tabularx}
\usepackage[usenames]{color}
\usepackage{xcolor}
\usepackage{float}
\usepackage{multirow}
\usepackage{colortbl}
\usepackage{setspace}
\usepackage{booktabs}
\usepackage{makecell}
\usepackage[normalem]{ulem}
\usepackage[most]{tcolorbox}

\usepackage[utf8]{inputenc}
\usepackage{graphicx}
\usepackage{subfig}
\usepackage{bm}
\usepackage{amssymb}

\usepackage{color}
\usepackage{textcomp}
\usepackage{listings}
\definecolor{codeblue}{rgb}{0,0,0.6}
\definecolor{codeviolet}{rgb}{0.39,0,0.55}
\definecolor{codegreen}{rgb}{0,0.6,0}
\definecolor{codegray}{rgb}{0.5,0.5,0.5}
\definecolor{codepurple}{rgb}{0.84,0,0.76}
\definecolor{codeback}{rgb}{0.9804,0.9882,0.9882}
\lstdefinelanguage{python}
{
    alsoletter={.},
    morekeywords={access,and,break,class,continue,def,del,elif,else,except,exec,finally,for,from,global,if,import,in,is,lambda,not,or,pass,print,raise,return,try,while},
keywords=[2]{abs,all,any,basestring,bin,bool,bytearray,callable,chr,classmethod,cmp,compile,complex,delattr,dict,dir,divmod,enumerate,eval,execfile,file,filter,float,format,frozenset,getattr,globals,hasattr,hash,help,hex,id,input,int,isinstance,issubclass,iter,len,list,locals,long,map,max,memoryview,min,next,object,oct,open,ord,pow,property,range,raw_input,reduce,reload,repr,reversed,round,set,setattr,slice,sorted,staticmethod,str,sum,super,tuple,type,unichr,unicode,vars,xrange,zip,apply,buffer,coerce,intern},
    sensitive=false,
    morecomment=[l]{\#},
    morestring=[b]"
}
\renewcommand{\lstlistingname}{\bfseries Listing}
\makeatletter
\def\fnum@lstlisting{%
  \lstlistingname
  \ifx\lst@@caption\@empty\else~\thelstlisting\normalfont\fi}%
\makeatother

\usepackage[all]{hypcap}
\usepackage{relsize}
\usepackage{fancyhdr}

\makeatletter
\def\p@subsection{}
\makeatother

\definecolor{Blue}{rgb}{0,0.31,0.75}
\definecolor{Red}{rgb}{1,0,0}

\DeclareMathOperator*{\Sum}{\mathlarger{\mathlarger{\sum}}}

\newcommand{\orcidicon}[1]{\href{https://orcid.org/#1}{\includegraphics[height=\fontcharht\font`\B]{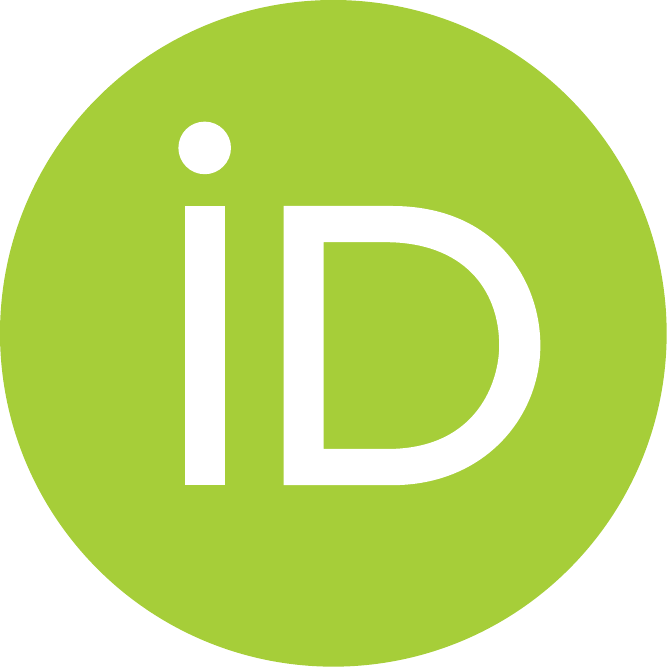}}}

\newcommand{\aTitle}{PyEOC: a Python Code for Determining Electro-Optic Coefficients of Thin-Film Materials}
\newcommand{\sTitle}{PyEOC: a Python Code for Determining Electro-Optic Coefficients of Thin-Film Materials}
\newcommand{\LMOPSUL}{Université de Lorraine, CentraleSupélec, LMOPS, F-57000 Metz, France.}
\hypersetup{
    bookmarks = true,
    pdfauthor = {Sidi {Ould Saad Hamady}},
    pdfsubject = {\sTitle},
    pdftitle = {\aTitle},
    pdfkeywords={PyEOC, Electro-Optic, Coefficient, Pockels, Modulation, Refractive Index, Thin Films, Semiconductors, Code, Python.},
    pdfproducer = {pdflatex},
    pdfcreator = {pdflatex},
    baseurl = http:article\_eoc.pdf,
    pdfdisplaydoctitle = true,
    pdfstartview = Fit,
    linktoc = all,
    colorlinks = true,
    urlcolor = black,
    linkcolor = black,
    citecolor = black
}
\newcolumntype{C}{>{\centering\arraybackslash}X}
\DeclareCaptionLabelFormat{LabelFigure}{\textbf{Figure #2}}
\DeclareCaptionLabelFormat{LabelTable}{\textbf{Table #2}}
\newcommand{\boldref}[1]{\textbf{\autoref{#1}}}

\graphicspath{{./}}

\begin{document}


\fancyhf{}
\fancyhead[LO]{S {Ould Saad Hamady} \hfill \sTitle}
\fancypagestyle{plain}{
\fancyfoot[C]{\thepage}}

\pagestyle{plain}

\title{\aTitle}

\author{Sidi {Ould Saad Hamady}\,\orcidicon{0000-0002-0480-6381} ~\\~\\ \LMOPSUL}

\setcounter{footnote}{0}
\maketitle

\begin{abstract}
    PyEOC is an open-source Python code for determining electro-optic (EO) coefficients of thin-film materials from the static and dynamic reflectivity measurements. It uses the experimental results, the transfer-matrix method and implements a robust fitting procedure to precisely calculate the EO coefficients. The developed code is applied to a Pt/SBN/Pt/MgO structure and can be easily adapted to any multilayer planar structure. The values of the EO coefficients determined using PyEOC are in excellent agreement with those obtained in the literature and this code will make it possible to explore EO properties of other thin-film materials, in particular III-V and III-N semiconductors. PyEOC is released under the permissive open-source MIT license. It is made available at \url{https://github.com/sidihamady/PyEOC} and depends only on standard Python packages (NumPy, SciPy and Matplotlib).
\end{abstract}



\section{Objectives and Methodology}
\label{section:objmethod}
\begin{figure}
\centering
   \includegraphics[width=0.5\linewidth]{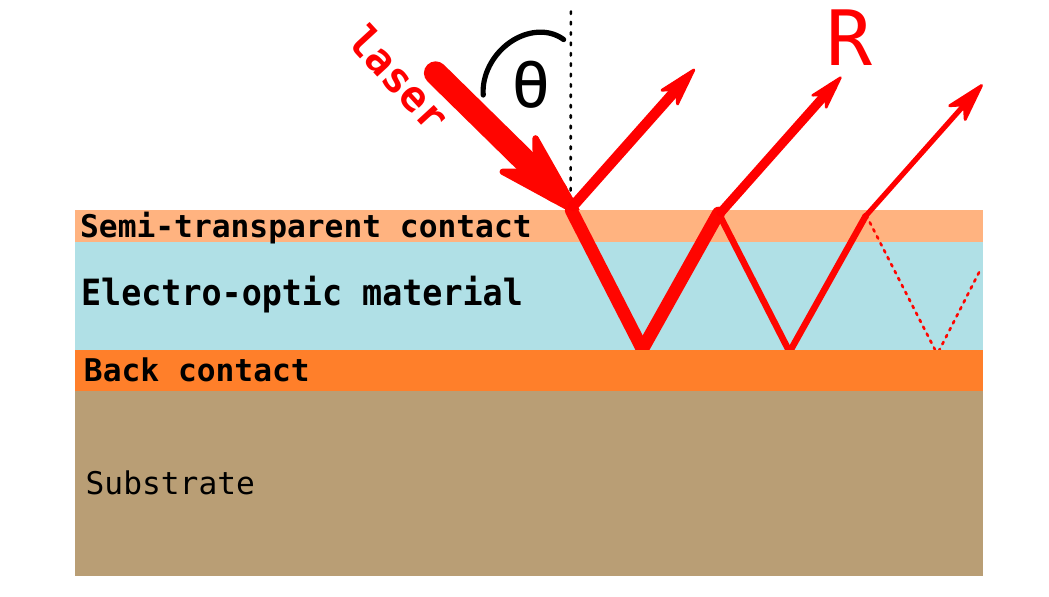}
   \caption{The standard multilayer structure used to measure the electro-optic (and converse piezoelectric) coefficients of a thin-film material. The film to characterize is sandwiched between two thin metallic contacts (\emph{e.g.} gold, platinum): a semi-transparent top contact and a thick bottom contact. The voltage is applied between these two contacts and the reflectance measured with respect to the incident angle $\mathrm{\Theta}$.}
   \label{fig:structure}
\end{figure}

Electro-optical (EO) materials are of vital interest in almost all areas of photonics and telecommunications. Their use is based on their optical properties with the linear modulation of the refractive index under the effect of an applied electric field. This is the Pockels effect, a detailed development of which can be found in reference \cite{saleh2019fundamentals}. The main used EO material is still lithium niobate $\mathrm{LiNbO_3}$ (LN) but various alternative materials have been developed, in particular thin films of $\mathrm{BaTiO_3}$ or $\mathrm{(Sr,Ba)Nb_2O_6}$ (SBN) \cite{hamze2020design,cuniot2011simultaneous} or semiconductors such as gallium arsenide (GaAs) \cite{sinatkas2021electro} or, to a lesser extent, gallium nitride (GaN) \cite{long1995gan,cuniot2014electro}. The precise and reliable determination of the electro-optical coefficients is therefore essential to characterize these materials, to optimize them and to integrate them into applications. This is particularly essential in the case of materials such as III-V semiconductors with EO coefficients that are very low compared to those of LN.
Cuniot-Ponsard \emph{et al.} proposed an experimental procedure to reliably measure the EO (and converse piezoelectric) coefficients using a Fabry-Perot interferometric setup with an example shown in \boldref{fig:structure}. In the present work, I implemented the first open-source code to analyze and extract the EO coefficients from the reflectivity measurements using this experimental procedure. 
    The variation $\mathrm{\Delta R}$ of the reflectivity $\mathrm{R}$ of such a multilayer planar structure (as shown in \boldref{fig:structure}) for the transverse electric (TE) polarization is given by:
    \begin{equation}
    \label{eqn:RTE}
    \Delta R_{TE} ~=~ \left( \frac{\delta R}{\delta n_o}  \right) \Delta n_o ~+~ \left( \frac{\delta R}{\delta k_o}  \right) \Delta k_o ~+~ \left( \frac{\delta R}{\delta d}  \right) \Delta d
    \end{equation}
Where $\mathrm{n_o}$, $\mathrm{k_o}$ and $\mathrm{d}$ are the ordinary refractive index, the ordinary extinction coefficient and the film thickness. $\mathrm{R}$ and $\mathrm{\Delta R}$ depend of course on the incident angle $\mathrm{\Theta}$.
For the transverse magnetic (TM) polarization, the expression is similar:
    \begin{equation}
    \label{eqn:RTM}
    \Delta R_{TM} ~=~ \left( \frac{\delta R}{\delta n_e}  \right) \Delta n_e ~+~ \left( \frac{\delta R}{\delta k_e}  \right) \Delta k_e ~+~ \left( \frac{\delta R}{\delta d}  \right) \Delta d
    \end{equation}
Where, similarly, $\mathrm{n_e}$, $\mathrm{k_e}$ are the extraordinary refractive index and the extraordinary extinction coefficient.

The experimental and analysis procedure is organized as follows:

\begin{itemize}
    \item   Step \#1: The static reflectivity $\mathrm{R}$ is measured for each value of the incident angle $\mathrm{\Theta}$ in both TE and TM polarizations.
    \item   Step \#2: The experimental $\mathrm{R(\Theta)}$ curves are numerically fitted using a theoretical procedure based on the transfer-matrix method with the Fresnel coefficients representing the coherent thin-film multilayers stack. The fitting procedure thus makes it possible to have the values of $\mathrm{n_x}$, $\mathrm{k_x}$ and $\mathrm{d}$ and to be able to calculate numerically the partial derivatives $\mathrm{\frac{\delta R}{\delta n_x}}$, $\mathrm{\frac{\delta R}{\delta k_x}}$ and $\mathrm{\frac{\delta R}{\delta d}}$ for each value of $\mathrm{\Theta}$ and for each polarization.
    \item  Step \#3: The dynamic (i.e. with the application of an AC voltage) reflectivity $\mathrm{\Delta R}$ is measured for each value of the incident angle $\mathrm{\Theta}$ in TE and TM polarizations.
    \item  Step \#4: Use (i) the measured $\mathrm{\Delta R (\Theta)}$ for three $\mathrm{\Theta}$ values; (ii) the derivatives calculated in step \#2; (iii) solve the obtained $\mathrm{3 \times 3}$ linear system (for the TE polarization, similar for the TM one) to obtain $\mathrm{\Delta n_x}$, $\mathrm{\Delta k_x}$ and $\mathrm{\Delta d}$:
    
    \begin{align}
    \begin{split}
    \Delta R_{TE}^{\Theta_1} ~&=~ \left( \frac{\delta R}{\delta n_o}  \right)^{\Theta_1} \Delta n_o ~+~ \left( \frac{\delta R}{\delta k_o}  \right)^{\Theta_1} \Delta k_o ~+~ \left( \frac{\delta R}{\delta d}  \right)^{\Theta_1} \Delta d \\
    \Delta R_{TE}^{\Theta_2} ~&=~ \left( \frac{\delta R}{\delta n_o}  \right)^{\Theta_2} \Delta n_o ~+~ \left( \frac{\delta R}{\delta k_o}  \right)^{\Theta_2} \Delta k_o ~+~ \left( \frac{\delta R}{\delta d}  \right)^{\Theta_2} \Delta d \\
    \Delta R_{TE}^{\Theta_3} ~&=~ \left( \frac{\delta R}{\delta n_o}  \right)^{\Theta_3} \Delta n_o ~+~ \left( \frac{\delta R}{\delta k_o}  \right)^{\Theta_3} \Delta k_o ~+~ \left( \frac{\delta R}{\delta d}  \right)^{\Theta_3} \Delta d
    \end{split}
    \end{align}

\end{itemize}

In step \#4, in theory only three values of $\mathrm{\Theta}$ are needed to obtain and solve the $\mathrm{3 \times 3}$ linear system. In practice, this step is much more trickier because the experimental values are sensitive to the precision and quality of the measurement as well as the data processing methodology. An iterative procedure, implemented in PyEOC, is thus necessary to extract reliably $\mathrm{\Delta n_x}$, $\mathrm{\Delta k_x}$ and $\mathrm{\Delta d}$ by using the whole experimental data and numerically fitting $\mathrm{\Delta R (\Theta)}$.

After obtaining $\mathrm{\Delta n_x}$, $\mathrm{\Delta k_x}$ and $\mathrm{\Delta d}$ ($x$ being $o$ for TE and $e$ for TM), the electro-optic ($\mathrm{r_{13}}$) and converse piezoelectric ($\mathrm{d_{33}}$) coefficients are calculated as follows:

\begin{align}
\begin{split}
r_{13} ~&=~ \frac{\Delta \left( \frac{1}{n_o^2}  \right)}{ \left( \frac{\Delta V}{d} \right)} ~=~  \frac{-2 d \Delta n_o}{n_o^3 \Delta V} \\
d_{33} ~&=~ \frac{\Delta d}{\Delta V} 
\end{split}
\end{align}
Where $\mathrm{n_o}$ is the ordinary refractive index, $\mathrm{\Delta V}$ the applied AC voltage amplitude, $\mathrm{d}$ the film thickness (thus $\mathrm{\Delta V / d}$ is the applied electric field).

\section{Software Implementation}
\label{section:software}
PyEOC is coded in vanilla Python and uses only standard packages \emph{e.g.} NumPy \cite{numpy2011}, SciPy \cite{scipy2020} and Matplotlib \cite{hunter2007matplotlib}. It uses object-oriented programming methodology to ensure better software quality and easier maintenance/update. It contains one class implementing the core functionalities (data input, calculations, fitting, consistency check, plotting and so on). The code is also easy to convert to the procedural approach if the latter is more suitable for a specific project. As usual, the basic programming rules were followed: (i) simplicity (\emph{keep it simple and straightforward}); (ii) modularity and easy maintenance; (iii) readability and code correctness.

The code is self-documented and easy to read/understand. Basically it consists of three blocks of methods:

\begin{itemize}
\item \texttt{Data input} methods, to read and filter the four experimental measurements (static reflectivity, dynamic reflectivity, both in TE and TM).

\item \texttt{Calculation/fitting} methods, implementing the algorithms detailed in \autoref{section:theory}. The transfer-matrix method part uses the Byrnes' tmm code \cite{byrnes2016multilayer}.

\item \texttt{Plotting/report} methods, to visualize the results and plot the reflectivity, derivatives, intensity variation in the structure, etc. 
\end{itemize}

PyEOC is available at \url{https://github.com/sidihamady/PyEOC} and can be used without any installation procedure. It is also packaged as a standard installable Python module (with {\ttfamily setup.py}). It includes an example detailed in the next section.
    

\section{An Example of Application}
\label{section:application}

PyEOC was used to analyze the experimental data of the structure studied by Cuniot-Ponsard \emph{et al.} \cite{cuniot2011simultaneous}. This structure consists of a $\mathrm{(Sr,Ba)Nb_2O_6}$ (SBN) thin film sandwiched between two platinum contacts, as illustrated in \boldref{fig:structure}. The used substrate is a bulk MgO with a thickness of $\mathrm{500~ \mu m}$. The used code is given in \boldref{code:example} and included in the distribution.   

The AC voltage amplitude is 1 V. The following values were given as starting values for the fitting algorithm:

\begin{itemize}
\item \texttt{Platinum contact}: thickness (top contact) = 22.6 nm; thickness (bottom contact) = 70 nm; complex index = 2.33 + 4.14j.

\item \texttt{SBN}: thickness = 754.5 nm; complex ordinary index = 2.3 + 0.0515j; complex extraordinary index = 2.26 + 0.0515j.

\item \texttt{MgO substrate}: thickness = 500000 nm; complex index = 1.7346 + 0.0j.

\end{itemize}

\begin{figure*}
\centering
   \includegraphics[width=1\linewidth]{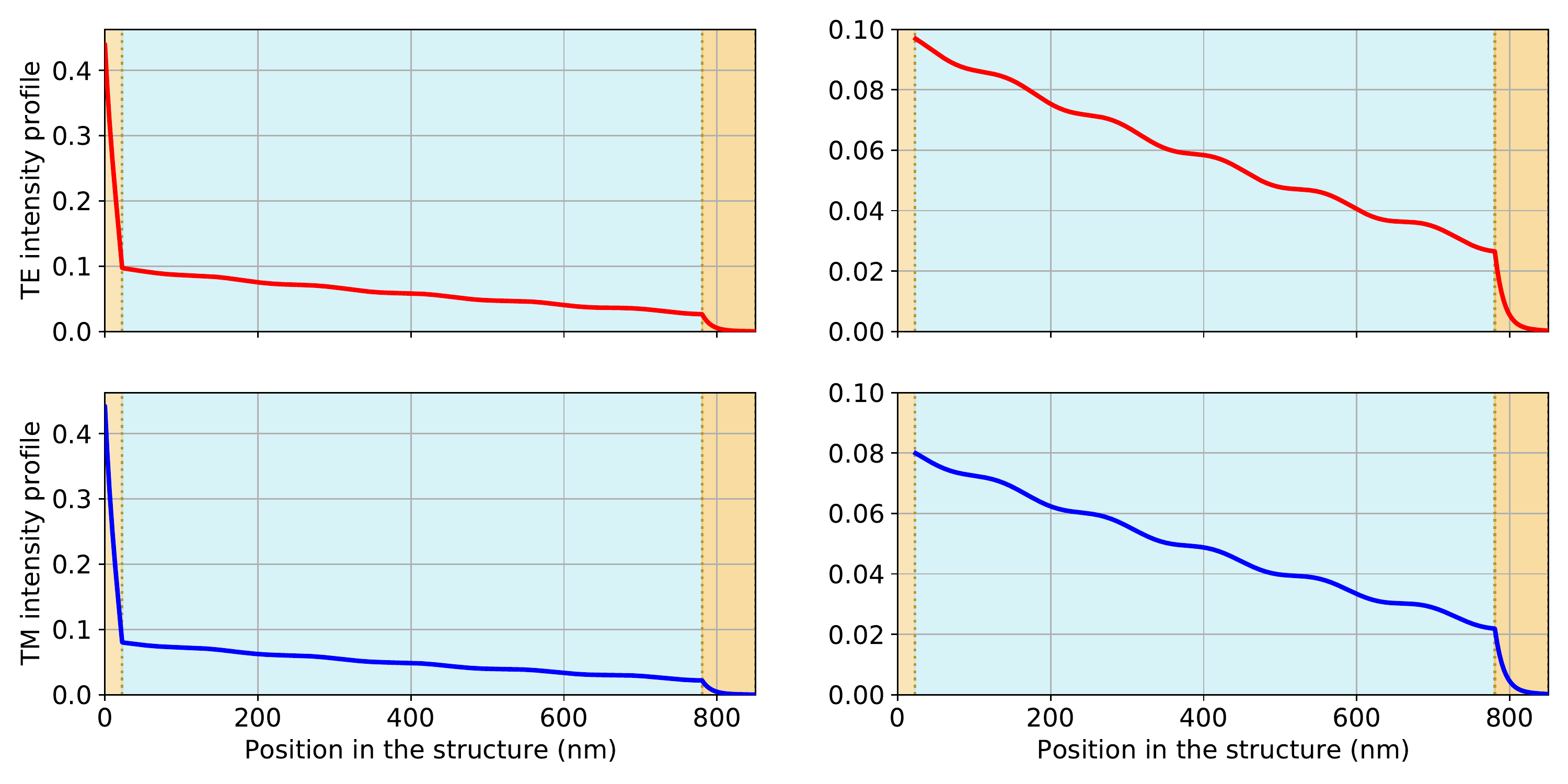}
   \caption{Intensity variation with the position in the Pt/SBN/Pt structure in both TE and TM polarization. The figures on the right represent a zoom in the active layer.}
   \label{fig:poynting}
\end{figure*}

\begin{figure*}
\centering
   \includegraphics[width=1\linewidth]{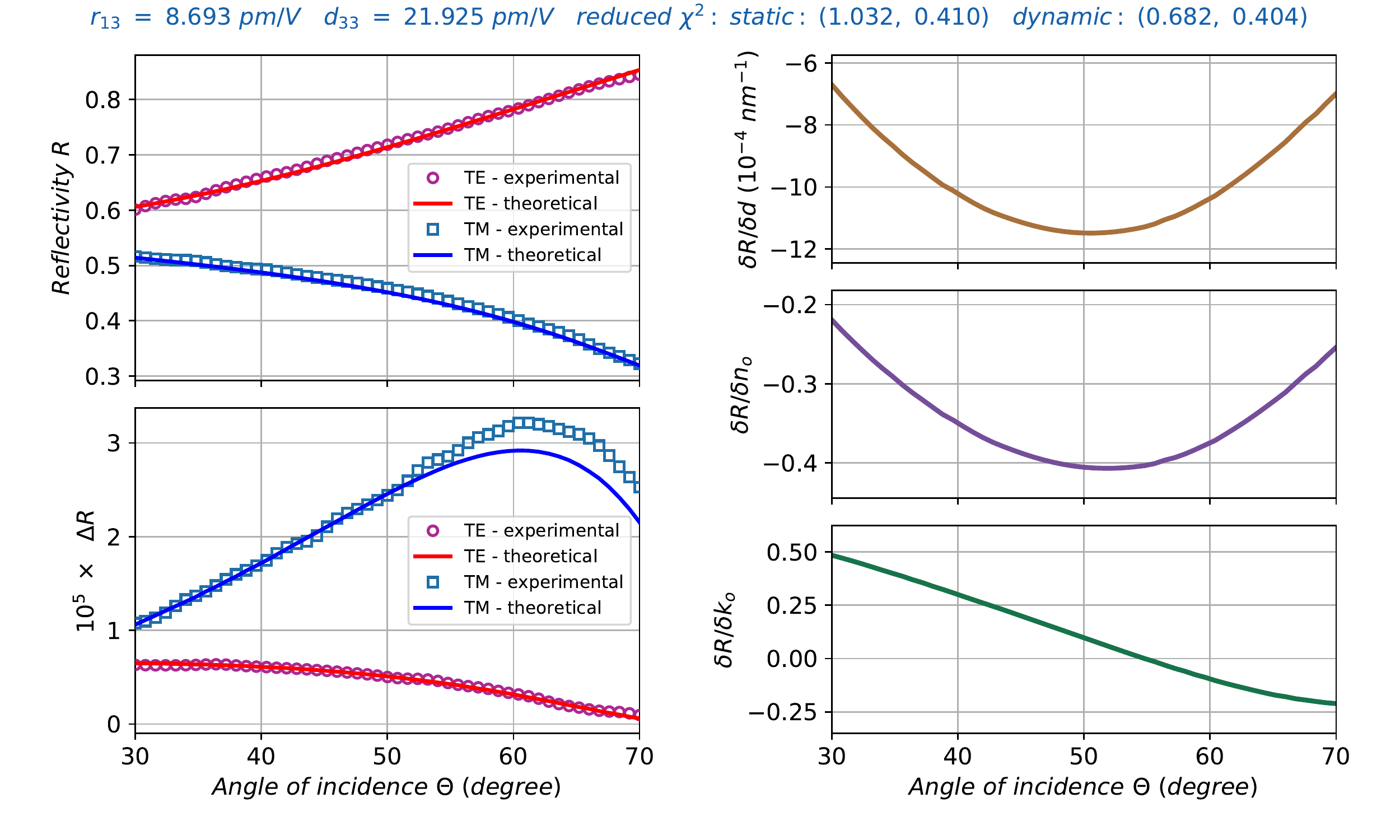}
   \caption{Experimental \cite{cuniot2011simultaneous} and theoretical calculated static reflectivity $\mathrm{R}$ (top-left), dynamic reflectivity $\mathrm{\Delta R}$ (bottom-left) and the calculated TE derivatives (right). The symbols indicate the experimental data curves and the lines indicate the fitted curves.}
   \label{fig:sbn}
\end{figure*}

\boldref{fig:poynting} shows the variation of the light intensity with respect to the position in the structure in both TE and TM polarization, and \boldref{fig:sbn} shows the experimental and the calculated static and dynamic reflectivity and the obtained derivatives. The obtained values of the  coefficients are:

\begin{align}
\begin{split}
r_{13} ~&=~ 8.69 ~pm~V^{-1} \\
d_{33} ~&=~ 21.92 ~pm~V^{-1} \\
\frac{\Delta k_{o}}{\Delta V} ~&=~ 11.10 \times 10^{-6} ~V^{-1}
\end{split}
\end{align}

To compare to the values obtained by Cuniot-Ponsard \emph{et al.} \cite{cuniot2011simultaneous}:

\begin{align}
\begin{split}
r_{13} ~&=~ 8.5~ \pm ~ 1.3 ~pm~V^{-1} \\
d_{33} ~&=~ 21~ \pm ~ 4 ~pm~V^{-1} \\
\frac{\Delta k_{o}}{\Delta V} ~&=~ (9.8 ~\pm~ 0.6) \times 10^{-6} ~V^{-1}
\end{split}
\end{align}

The agreement is thus excellent considering the given precision. The accuracy of the fitting procedure is estimated using the reduced $\mathrm{\chi^2}$ value:
\begin{align}
\begin{split}
\chi^2 ~&=~ \frac{1}{N ~-~ p} \Sum_{\Theta}{ \left( \frac{R_\Theta^{exp} ~-~ R_\Theta^{th}}{\sigma_\Theta}  \right)^2 }
\end{split}
\end{align}
Where $\mathrm{R_\Theta^{exp}}$ is the experimental reflectivity at a given incident angle $\mathrm{\Theta}$ and $\mathrm{R_\Theta^{th}}$ the theoretical calculated one. $\mathrm{N}$ is the number of measured values and $\mathrm{p}$ is the number of fitting parameters (thus $\mathrm{N ~-~ p}$ is the degree of freedom). $\mathrm{\sigma_\Theta}$ is the standard deviation representing the uncertainty on the measured data. In the example taken in this article, $\mathrm{\sigma_\Theta}$ was chosen at $\mathrm{0.5 \%}$ of the measured value for the static reflectivity (both in TE and TM polarization) and $\mathrm{10 \%}$ for the dynamic reflectivity. These values should be adapted \emph{e.g.} to take into account the variation in the measurement accuracy with the incident angle and, of course, with the used setup. If we consider that the measured data are reliable within the given uncertainty and the used optical model is accurate then the value of the reduced $\mathrm{\chi^2}$ should be relatively close to the unity. This value is therefore a useful indicator in this tricky procedure.


\section{Conclusion}
In this article, I have presented PyEOC, an open-source Python package for determination of the electro-optic coefficients from experimental reflectivity data. The code was developed in such a way as to ensure its evolution and maintenance. It was applied to a SBN structure with an excellent agreement with the published values. A future evolution will concern: (i) the adaptation of the code for other optical materials and the reliable extraction of other electro-optic coefficients \emph{e.g.} $\mathrm{r_{33}}$; (ii) the application to the determination of electro-optic coefficients of semiconductor thin-film materials for new applications in optics.


\begin{lstlisting}[linewidth=1\columnwidth,language=python,label=code:example,caption={Example of using PyEOC to analyze a SBN structure experimentally studied by Cuniot-Ponsard \emph{et al.} \cite{cuniot2011simultaneous}.}]
# -*- coding: utf-8 -*-

import sys, os
#sys.path.insert(0, "/path/to/PyEOC")
from PyEOC import *     # import PyEOC core class

Structure = PyEOC(
    'SBN',  # structure name included in the PyEOC class
    # measurement data: static  reflectivity vs angle (TE and TM)
    #                   dynamic reflectivity vs angle (TE and TM)
    #                   four files with tab-separated columns
    # 'SBN' data extracted from Cuniot-Ponsard et al. in JAP 109, 014107 (2011)
    'SBN_Reflectivity_TE.txt',      'SBN_Reflectivity_TM.txt',
    'SBN_Reflectivity_Dyn_TE.txt',  'SBN_Reflectivity_Dyn_TM.txt'
)

Structure.wavelength = 633  # laser wavelength in nm
Structure.voltage = 1.0     # applied voltage amplitude in volts

# the incident angle theta starting three values and range
# the choosen theta values should correspond to a "smooth" and "different" part of the DR and delta_R / delta_? data
Structure.theta_manual  = [35.0 * Structure.toradian, 40.0 * Structure.toradian, 45.0 * Structure.toradian]
Structure.thetaDelta    =   2.0 * Structure.toradian
Structure.thetaStart    =  30.0 * Structure.toradian
Structure.thetaEnd      =  70.0 * Structure.toradian

Structure.thickness[1] =  22.6  # Pt thickness (nm)
Structure.thickness[2] = 758    # SBN thickness (nm)
Structure.refractiveindexo[2] = 2.30 + 0.0515624j
Structure.refractiveindexe[2] = 2.26 + 0.0515624j

Structure.fit_dynamic = True    # fit the dynamic reflectivity?
Structure.fit(report = True)    # start fitting and report
Structure.plot()                # plot the fitted curves
#Structure.plotpoynting()       # plot the intensity profile

\end{lstlisting}


\section*{Conflict of Interests}
The author declares that he has no conflicts of interest.


\bibliography{article_eoc}


\end{document}